\documentclass[structabstract,openany]{aa}  

\usepackage[dvipdf]{graphicx,graphics}
\usepackage{lscape}
\usepackage{color}
\usepackage{txfonts}
\usepackage{natbib}
\bibpunct{(}{)}{;}{a}{}{,} 

\usepackage{marvosym}
\usepackage{color,hyperref}
\definecolor{darkblue}{rgb}{0.0,0.0,0.3}
\hypersetup{colorlinks,breaklinks,
            linkcolor=blue,urlcolor=blue,
            anchorcolor=blue,citecolor=blue}

\begin{document}

\newcommand{\kms}{km s$^{-1}$}

\title{Extended warm and dense gas towards W49A: starburst conditions in our Galaxy?}
\titlerunning{Extended warm and dense gas towards W49A}

   \author{Z. Nagy
          \inst{1,2},
          F. F. S. van der Tak\inst{2,1},
          G. A. Fuller\inst{3},
          M. Spaans\inst{1}
          \and
		  R. Plume\inst{4}
          }
          
\authorrunning{Z. Nagy et al.}          

\institute
{Kapteyn Astronomical Institute, University of Groningen, P.O. Box 800, 9700AV, Groningen, The Netherlands \\
\email{nagy@astro.rug.nl} 
\and
SRON Netherlands Institute for Space Research, P.O. Box 800, 9700AV, Groningen, The Netherlands
\and
Jodrell Bank Centre for Astrophysics, School of Physics and Astronomy, University of Manchester, Manchester, M13 9PL, UK
\and
Department of Physics and Astronomy, University of Calgary, Calgary, T2N 1N4, AB, Canada
}

\date{Received December 5, 2011; accepted March 28, 2012}
 
  \abstract
  { The star formation rates in starburst galaxies are orders of
    magnitude higher than in local star-forming regions, and the
    origin of this difference is not well understood.}
  { We use sub-mm spectral line maps to characterize the physical conditions
    of the molecular gas in the luminous Galactic star-forming region W49A and compare them with the conditions in
    starburst galaxies.}
  { We probe the temperature and density structure of W49A using
    H$_2$CO and HCN line ratios over a $2'\times2'$ ($6.6\times6.6$
    pc) field with an angular resolution of 15$''$ ($\sim$0.8 pc)
    provided by the JCMT Spectral Legacy Survey.  We analyze the
    rotation diagrams of lines with multiple transitions with
    corrections for optical depth and beam dilution, and estimate
    excitation temperatures and column densities.}
%
  { Comparing the observed line intensity ratios with non-LTE radiative
    transfer models, our
    results reveal an extended region (about $1'\times1'$, equivalent to
    $\sim$3$~\times~$3 pc at the distance of W49A) of warm ($>$100 K) and dense
    ($>$10$^5$ cm$^{-3}$) molecular gas, with a mass of $2\times10^4-2\times10^5~M_\odot$ (by applying abundances derived 	
    for other regions of massive star-formation).
    These temperatures and densities in W49A are comparable to those
    found in clouds near the center of the Milky Way and in starburst
    galaxies.  }
%
  { The highly excited gas is likely to be heated via shocks from the stellar
    winds of embedded, O-type stars or alternatively due to UV irradiation, or
    possibly a combination of these two processes. Cosmic rays, X-ray irradiation
    and gas-grain collisional heating are less likely to be the source of the heating in the case of
    W49A.  }

\maketitle

\section{Introduction}

The formation of stars proceeds on scales ranging from 1000 AU-sized isolated
low-mass cores to $\sim$10 pc-sized GMCs with embedded massive clusters. The
observed levels of star-formation activity also vary greatly: from
$\sim$10$^{-4}~M_\odot/{\rm{yr}}$ in dwarf galaxies like I Zw 18
\citep{aloisi1999} to $\sim$3 $M_\odot/{\rm{yr}}$ in the Milky Way to
$\sim$10$^3~M_\odot/{\rm{yr}}$ in starburst galaxies
(e.g. \citealp{solomonvandenbout2005}). The origin of this large spread in
star formation rate may lie in different physical conditions with the gas
forming the stars, the nature and role of feedback effects or other
mechanisms.  As one of the the most active star-forming regions in the Galactic disk, W49A
provides a relatively nearby laboratory in which to study intense star
formation activity and may provide insights into explaining the large observed
range in star-formation activity in galaxies.
 	    
W49A is a luminous ($>$10$^7~L_\odot$) and massive ($\sim$10$^6~M_\odot$) star-forming region \citep{sievers1991} at a distance of 11.4 kpc \citep{gwinn1992}. W49A contains several signatures of high star-formation activity, in particular, strong H$_2$O maser emission \citep{genzel1978}.
Line ratios of HNC, HCN and HCO$^+$ and their isotopologues suggest
similar conditions to starburst galaxies \citep{roberts2011}.
Several ideas have been proposed to explain the high star-formation
activity of W49A.

\cite{welch1987} propose a large-scale inside-out gravitational
collapse by interpreting two velocity components of HCO$^+$ J=1$-$0 spectra
measured toward the ring-like configuration of UC H{\sc ii} regions in
terms of an infall profile.
\cite{serabyn1993} interpret the two-components of the double-peak line
profile seen in CS (J=10$-$9, 7$-$6, 5$-$4, 3$-$2) and C$^{34}$S (J=10$-$9, 7$-$6, 5$-$4) transitions as coming from different clouds, and suggest
that massive star formation in W49A is triggered by a large-scale cloud
merger.
\cite{williams2004} argue that one cannot distinguish
between a global collapse model of a large cloud and a multiple-cloud model,
based on HCO$^+$ (J=3$-$2 and 1$-$0) and C$^{18}$O (J=2$-$1) data. \citet{roberts2011}
observe single-peaked HC$^{18}$O$^+$ (J=4$-$3) line profiles, which supports the
large-scale infall hypothesis.
In $^{13}$CO and C$^{18}$O J=2$-$1 maps of the W49A region \cite{peng2010} find evidence of expanding
shells with the dense clumps in the region lying peripherally along the
shells. These expanding shells could provide a natural explanation of the
observed molecular line profiles, and triggered massive star formation in
W49A.

This paper reports new kinetic temperature and density estimates for a
$2'\times2'$ region of W49A based on single dish, JCMT images with 15$''$
resolution. These indicate the presence of extended warm and dense gas towards
the central region of the source. We use these results to compare W49A to
starburst galaxies where similar results have been reported.

\section{Observations and Data reduction}

This paper presents results from data obtained as part of the JCMT
Spectral Legacy Survey (SLS, \citealp{plume2007}) covering between 330
and 373 GHz which has been carried out at the James Clerk Maxwell
Telescope (JCMT)\footnote{The James Clerk Maxwell Telescope is
  operated by The Joint Astronomy Centre on behalf of the Science and
  Technology Facilities Council of the United Kingdom, the Netherlands
  Organisation for Scientific Research, and the National Research
  Council of Canada.} on Mauna Kea, Hawai'i. The data were taken using
the HARP-B receiver and the ACSIS correlator
\citep{buckle2009}. 
HARP-B consists of 16 pixels or receptors separated by 30 arcsec with a foot print of 2 arcmin.

The observations were carried out using jiggle -
position switch mode. Each map covers a $2'\times2'$ field sampled
every 7.5$''$, half of the JCMT beam width at the relevant frequencies
(15$''$ at 345 GHz). The spectral resolution at these frequencies is
$\sim$0.8 \kms and the beam efficiency is 0.63 \citep{buckle2009}.
The spectra were taken using an off-position 14$'$ to the northeast of
the source.

Data reduction was done manually using the tools in the 
\texttt{Starlink} software package. 
The raw time series files are processed in 1 GHz wide chunks. These files were
inspected for bad receptors, baselines and frequency spikes. Bad receptors,
bad baselines and spikes were masked in the time series files before they have
been converted to three-dimensional data cubes
(RA$\times$Dec$\times$frequency). These 3D cubes then were baseline corrected
by fitting and removing a linear or second-order polynomial baseline. The
baseline corrected cubes are then mosaicked to a larger cube covering an area
of $2'\times2'$ and the frequency range of interest. Single sideband system
temperatures for these observations are in the range between 255 K (at 360 GHz) and 700 K (at 373 GHz). The
analysis here primarily focuses on a number of spectral lines in the
360$-$373~GHz range plus a few selected lines from the 330$-$360 GHz part of the
survey. The whole, systematic data analysis of the full SLS frequency range is
the focus of a future paper.

We also use ancillary observations of the HCN $J$ = 3$-$2 transition made using JCMT
raster maps with half-beamwidth spacing using the Receiver A (211$-$276 GHz;
beamwidth $20''$; main beam efficiency: 0.69). The HCN 3$-$2 map covers approximately the same field that of
the SLS HARP maps. The HCN 3$-$2 data were reduced and calibrated with standard
Starlink procedures. Linear baselines were removed and the data were Hanning
smoothed. The map is a grid of 15 pixel $\times$ 15 pixels, where each pixel
covers 8$''\times$ 8$''$.

\begin{table}
\begin{minipage}[t]{\linewidth}\centering  
\caption{Summary of the molecular lines used in this paper}             
\label{table:ext_emission}      
  
\renewcommand{\footnoterule}{}                       
\begin{tabular}{l l c l c} 
\hline\hline        
Molecule&       Transition&                 Frequency&   $E_\mathrm{up}$&  rms ($T_\mathrm{mb}$)
\footnote{noise level measured over the
central $1'\times1.5'$, for 1 \kms~ channels.}\\ 
    &                     &                 (MHz)&       (K)&              (K)                   \\ 
\hline
\\[-1.2ex]
HCN&            $J$ = 3$-$2&                    265886.2&    25.5&             0.90\\
HCN&            $J$ = 4$-$3&                    354505.5&    42.5&             0.05\\
\hline
\\[-1.2ex]
H$_2$CO&        $5_{1,5}-4_{1,4}$&            351768.6&    62.5&             0.12\\
H$_2$CO&        5$_{0,5}-4_{0,4}$&            362736.1&    52.3&             0.05\\
H$_2$CO&        $5_{2,4}-4_{2,3}$&          363945.9&    99.5&             0.06\\ 
H$_2$CO&        $5_{4,1}-4_{4,0}$&          364103.3&    240.7&            0.06\\
H$_2$CO&        $5_{3,3}-4_{3,2}$&            364275.1&    158.4&            0.08\\      
H$_2$CO&        $5_{3,2}-4_{3,1}$&          364288.9&    158.4&            0.08\\  
H$_2$CO&        $5_{2,3}-4_{2,2}$&          365363.4&    99.7&             0.06\\
\hline
\\[-1.2ex]
CH$_3$OH&       $11_{0,11}-10_{1,9}$&       360848.9&    166.0&            0.04\\
CH$_3$OH&       $8_{1,7}-7_{2,5}$&          361852.3& 	 104.6&            0.05\\
CH$_3$OH&       $16_{1,15}-16_{0,16}$&      363440.3& 	 332.6&            0.06\\
CH$_3$OH&       $7_{2,5}-6_{1,5}$&          363739.8&    87.3&             0.06\\
CH$_3$OH&       $17_{1,16}-17_{0,17}$&      371415.5& 	 372.4&            0.09\\ 
\hline
\\[-1.2ex]
SO$_2$&         $34_{5,29}-34_{4,30}$&      360290.4& 	 611.9&            0.06\\ 
SO$_2$&         $20_{8,12}-21_{7,15}$&      360721.8& 	 349.8&            0.04\\
SO$_2$&         $21_{4,18}-21_{3,19}$&      363159.3&    252.1&            0.05\\ 
SO$_2$&         $24_{1,23}-24_{0,24}$&      363890.9&    280.5&            0.06\\
SO$_2$&         $23_{2,22}-23_{1,23}$&      363925.8& 	 259.9&            0.06\\
SO$_2$&         $25_{9,17}-26_{8,18}$&      364950.1&    497.1&            0.06\\
SO$_2$&         $15_{2,14}-14_{1,13}$&      366214.5&    119.3&            0.09\\   
SO$_2$&         $9_{6,4}-10_{5,5}$&         370108.6&    129.7&            0.11\\ 
SO$_2$&         $6_{3,3}-5_{2,4}$&          371172.5&    41.4&             0.09\\  
\hline     
                      
\end{tabular}
\end{minipage}
\end{table}		      

\section{Results}

\subsection{Line selection}

	 \begin{figure*}[!ht]
       \includegraphics[width=19 cm,trim=1cm 0 0 0,clip=true]{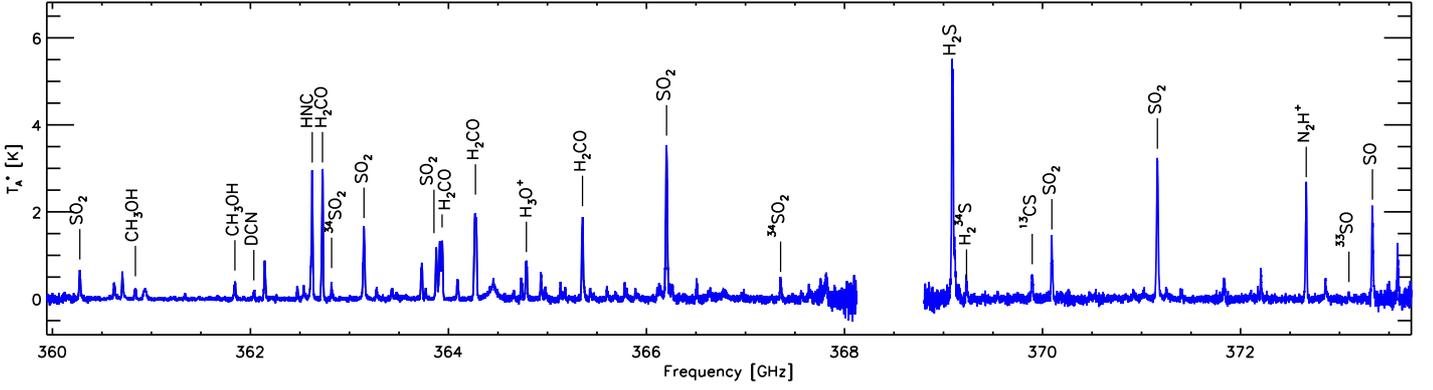}
       \caption{The spectrum towards the source center (RA(J2000)=19:10:13.4;
         Dec(J2000)=09:06:14) in the 360$-$373 GHz frequency range of the
         SLS. The main detected lines are labeled. The gap in the spectrum
         around 368.5 GHz is a region of high atmospheric extinction and was
         not observed.}
         \label{W49_spec}
     \end{figure*}

Figure \ref{W49_spec} shows the 360$-$373 GHz spectrum towards the central
     position (RA(J2000)=19:10:13.4; Dec(J2000)=09:06:14). More than 80 lines
     have been identified in this spectrum.  

     Our initial selection of lines for the analysis here is based on
     molecules that show multiple transitions in the high-frequency range
     (360$-$373 GHz) of the SLS, which include six H$_2$CO, five CH$_3$OH and
     nine SO$_2$ transitions (Table \ref{table:ext_emission}). These are used
     to trace the excitation conditions. One more H$_2$CO transition from the
     lower frequency part of the survey (H$_2$CO $5_{15}-4_{14}$) was added to
     be used as a tracer of kinetic temperatures, together with the
     $5_{33}-4_{32}$ transition from the high-frequency part of the SLS. The
     HCN 4$-$3 transition was selected from the lower frequency range, to
     trace the density structure, together with the HCN 3$-$2 transition.

\begin{center}
	 \begin{figure}[!hhh]
	   \centering
	   \includegraphics[width=9 cm,trim=0cm 0cm 7cm -0.5cm,clip=true]{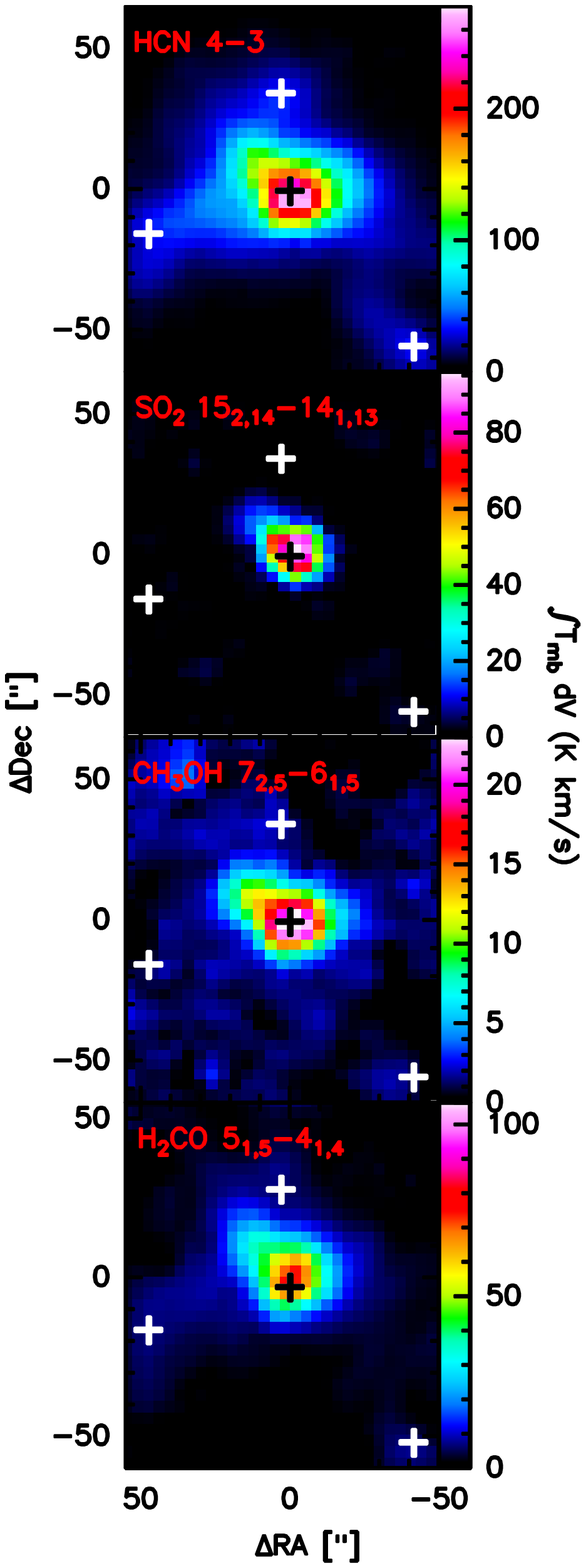}       
      \caption{The integrated intensity distribution of HCN 4$-$3, SO$_2$ $15_{2,14}-14_{1,13}$, CH$_3$OH $7_{2,5}-6_{1,5}$ and H$_2$CO $5_{15}-4_{14}$. The central position is shown with black plus signs, the off-center regions (Northern clump, Eastern tail and South-west clump) are also shown with white plus signs.}
         \label{hcn_regions}
     \end{figure}
    \end{center}

     In this paper, we restrict our analysis to our line selection summarized
     above and in Table~\ref{table:ext_emission}. A subsequent paper (Nagy et
     al, in prep.) will give more details on the chemical inventory including the full SLS
     frequency range (330$-$373 GHz) and on kinematics traced by various species with significant spatial extension.
Figure \ref{hcn_regions} shows the main regions around the center of W49A
     overplotted on the integrated intensity maps in the HCN 4$-$3, SO$_2$ 
     $15_{2,14}-14_{1,13}$, CH$_3$OH $7_{2,5}-6_{1,5}$ and H$_2$CO $5_{15}-4_{14}$ lines. The integrated intensities of the HCN and H$_2$CO lines show a similar distribution, indicating that their emission originates in the same volume. This is also supported by the line profiles of these two species (Nagy et al, in prep). 
     In the following sections, we will refer to these regions with the coordinates
     indicated in the figure: source center (RA(J2000) = 19:10:13.4;
     Dec(J2000) = 09:06:14), Eastern tail (RA(J2000) = 19:10:16.6; Dec(J2000)
     = 09:05:48), Northern clump (RA(J2000) = 19:10:13.6; Dec(J2000) =
     09:06:48) and South-West clump (RA(J2000)=19:10:10.6;
     Dec(J2000) = 09:05:18).

\subsection{Excitation conditions}

A well known method for the estimation of excitation temperatures and column
densities is the rotational diagram method (e.g. \citealp{turner1991}). It can
be applied to molecules with multiple observed transitions using three
assumptions: 1) the lines are optically thin, 2) the level populations can be
characterized by a single excitation temperature ('rotational temperature',
$T_\mathrm{rot}$) and 3) the emission is homogeneous and fills the telescope
beam. Then the measured integrated main-beam temperatures of lines ($\int
T_{\mathrm{MB}} \mathrm{d}V$ K km s$^{-1}$) is related to the column densities
of the molecules in the upper level ($N_\mathrm{u}$) by:
\begin{equation}
	\label{rotdiagram}
		\frac{N_\mathrm{u}}{g_\mathrm{u}}=\frac{N_{\mathrm{tot}}}{Q(T_{\mathrm{rot}})} \exp\left({-\frac{E_\mathrm{u}}{T_{\mathrm{rot}}}}\right)=\frac{1.67 \cdot 10^{14}}{\nu \mu^2 S} \int T_{\mathrm{MB}} \mathrm{d}V,
\end{equation}
with $g_\mathrm{u}$ the statistical weight of level u, $N_{\mathrm{tot}}$ the
total column density in cm$^{-2}$, $Q(T_{\mathrm{rot}})$ the partition
function for $T_{\mathrm{rot}}$, $E_\mathrm{u}$ the upper level energy in K,
$\nu$ the frequency in GHz, $\mu$ the permanent dipole moment in Debye and $S$
the line strength value. A linear fit to $\ln(N_\mathrm{u}/g_\mathrm{u})$ -
$E_\mathrm{u}$ gives $T_\mathrm{rot}$ as the inverse of the slope, and
$N_\mathrm{tot}$, the column density can be derived. The rotational
temperature would be expected to be equal to the kinetic temperatures if all
levels were thermalized.

\begin{figure}[!hh]
		\centering
		\includegraphics[width=9.5 cm]{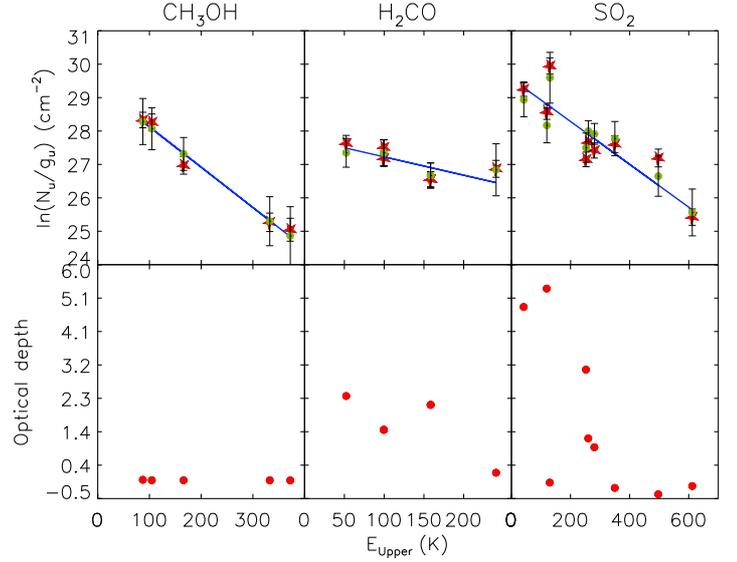}
    \caption{The results of the rotation diagram analysis for H$_2$CO toward the center. \emph{Top panels}: 
The results of the rotation diagram analysis before (red symbols) and after (green symbols) corrections for optical depth and beam dilution. The overplotted blue line corresponds to a linear fit to the rotational diagram without corrections. \emph{Bottom panels}: The corresponding best fit optical depths from the $\chi^2$ minimization.}			
		\label{popdiagram}
		\end{figure}

In the case of non-LTE excitation, a different excitation temperature
may characterize the population of each level relative to that of the
ground state or relative to that of any other level. In this case, the
three assumptions mentioned for the rotation diagram method do not
apply. Equation~\ref{rotdiagram} can be modified to include the effect
of optical depth $\tau$ through the factor
C$_\tau$=$\tau$/(1-e$^{-\tau}$) and beam dilution $f$
($=\Omega_s/\Omega_a$) \citep{goldsmith1999}, with $\Omega_s$ the size
of the emission region and $\Omega_a$ the size of the telescope beam.
\begin{equation}
\label{PD}
\ln \left(\frac{N_\mathrm{u}}{g_\mathrm{u}}\right)=\ln \left( \frac{N_{\mathrm{tot,1}}}{Q(T_{\mathrm{rot}})} \right) - \frac{E_\mathrm{u}}{k T_{\mathrm{ex}}} + \ln{(f)} - \ln{(C_\tau)}.
\end{equation}
According to equation \ref{PD}, for a given upper level,
$N^{\mathrm{obs}}_\mathrm{u}$ can be evaluated from a set of
$N_{\mathrm{tot,1}}$, $T_{\mathrm{ex}}$, $f$ and $C_\tau$. Since $C_\tau$ is a
function of $N_{\mathrm{tot,1}}$ and $T_{\mathrm{ex}}$, the independent
parameters are therefore $N_{\mathrm{tot,1}}$, $T_{\mathrm{ex}}$ and $f$.
Solving the equation for a 'source size' between 1$''$ and 15$''$ (which is
equivalent with a beam dilution factor between (1/15)$^2$ and uniform beam
filling), $\chi^2$ minimization gives $N_{\mathrm{tot,1}}$ and
$T_{\mathrm{ex}}$.
Figure \ref{popdiagram} shows the rotation diagrams for CH$_3$OH, H$_2$CO \&
SO$_2$ toward the central position with (green symbols) and without (red
symbols) taking into account the optical depth and beam dilution. Table
\ref{table:1} includes the parameters of the rotation diagram fit toward the
central position.

The CH$_3$OH lines are found to be optically thin ($\tau=0.004-0.02$), so the
rotation diagram fit is a good approximation to the column densities and 
excitation temperatures. The derived rotation temperature
is 83$\pm$7 K. The H$_2$CO and SO$_2$ lines are mostly optically thick with
optical depths in the range of 0.2$-$2.35 for H$_2$CO and 0.1$-$5.4 for
SO$_2$. The best fit excitation temperatures are 138 K for H$_2$CO and 115 K
for SO$_2$. The best fit total column densities are
$4.3\times10^{16}~{\rm{cm}}^{-2}$ for H$_2$CO and
$1.4\times10^{18}~{\rm{cm}}^{-2}$ for SO$_2$. The $\chi^2$ minimization for
H$_2$CO and SO$_2$ results in a best fit `source size' of $\sim$2$-$3$''$,
which is equivalent with 0.11$-$0.17 pc for W49A. This is consistent with the
sizes of hot cores and UC H{\sc{ii}} regions resolved by sub-arcsecond
resolution measurements, such as \citet{wilner2001},
\citet{depree1997,depree2004}.

\begin{figure}[!th]
	\centering
	\includegraphics[width=9.5 cm]{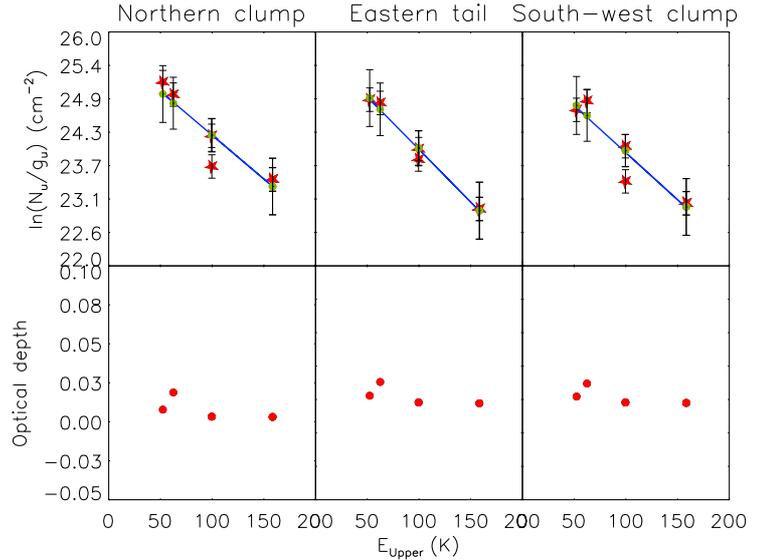}
    \caption{The results of the rotation diagram analysis for H$_2$CO at off-center positions. \emph{Top panels}: 
The results of the rotation diagram analysis before (red symbols) and after (green symbols) corrections for optical depth and beam dilution. The overplotted blue line corresponds to a linear fit to the rotational diagram without corrections. \emph{Bottom panels}: The corresponding best fit optical depths from the $\chi^2$ minimization.}		
	\label{popdiagram_regions}
	\end{figure}

Figure \ref{popdiagram_regions} shows the rotation diagrams for H$_2$CO toward the three main regions outside the center. 
The rotation diagram with corrections for optical depth and beam dilution (Figure \ref{popdiagram_regions}. and Table~\ref{table:2}.) gives similar excitation conditions toward the three main sub-regions (excitation temperatures: 67 K for the Northern clump, 56 K for the Eastern tail and 61 K for the South-west clump). These excitation temperatures are about half of the excitation temperature towards the central position (138 K). Since the lines are optically thin, the rotation temperatures are good estimates for the excitation temperatures. The best fit column densities are also similar between the off-center positions, about 1.7$-$2.2$\times10^{13}~{\rm{cm}}^{-3}$, almost three orders of magnitude less than the best fit column density towards the center. The data at these positions are consistent with optically thin emission and uniform beam filling.

\subsection{Kinetic temperature estimates}
\label{sect_kintemp}

To estimate the local physical conditions in W49A, we use ratios of lines from the same molecular species.
The kinetic temperature can be probed by the line ratios of H$_2$CO lines from the same $J$-state but different $K$-states. In this case, we use the o$-$H$_2$CO 5$_{15}-4_{14}$ (351.769 GHz) and o$-$H$_2$CO 5$_{33}-4_{32}$ (364.275 GHz) transitions, since they have been detected in a large part of the SLS field and have been found to be excellent tracers of kinetic temperature (Figure~\ref{line_ratios_H2CO}). Our calculations use the non-LTE radiative transfer program Radex \citep{vandertak2007}, molecular data file from the LAMDA database \citep{schoier2005} and H$_2$CO collision data from \citet{green1991}, which have been scaled by 1.37 to make a first order approximation for collisions with H$_2$. Uncertainties introduced by using these rates may be up to 50\% for the column densities, but less affect the densities and temperatures, since those are derived from line intensity ratios. More recent calculations by \citet{troscompt2009} are more accurate, however, include a more limited range of temperatures and transitions, therefore, were not used in our calculations.

Figure \ref{line_ratios_H2CO} shows the calculated line ratios as a function of kinetic temperature and H$_2$ density.
In the central $40''\times40''$ region, FWHM=12 km s$^{-1}$ was used as an initial parameter for Radex, while outside of this region we used FWHM=6 km s$^{-1}$, based on the observed values. However, there is no strong dependence on the FWHM of the lines. 
We apply a molecular column density of $N$(H$_2$CO)=$5\times10^{14}$ cm$^{-2}$ as an 'average' value for all the positions over the SLS field. This is slightly below the value that can reproduce the optical depths derived from rotation diagram method for the central position, assuming uniform beam filling.

\begin{figure*}[!th]
\includegraphics[width=9.5 cm,trim=1cm 0 0 9cm,clip=true]{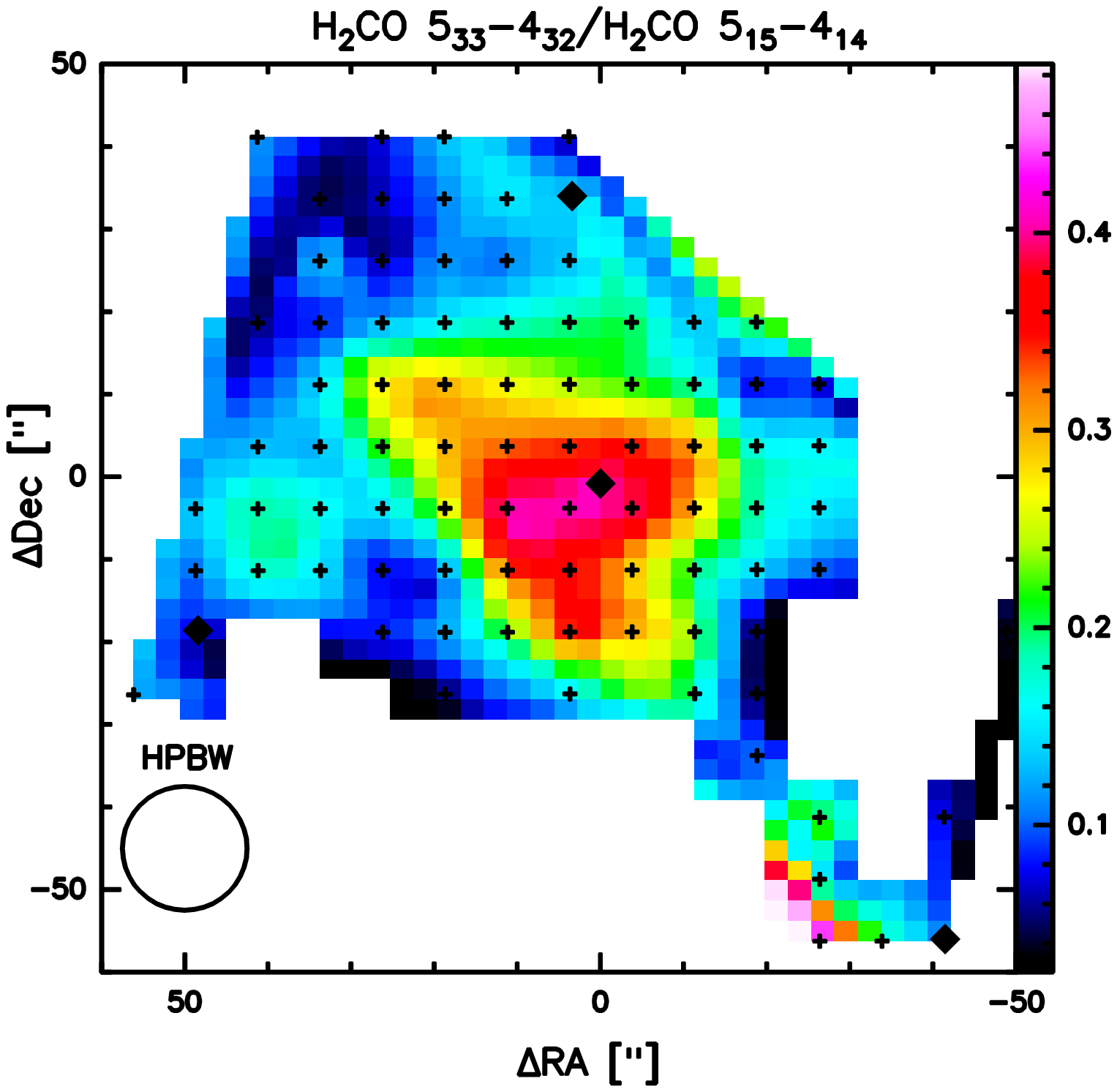}
\includegraphics[width=9.5 cm,trim=1cm 0 0 9cm,clip=true]{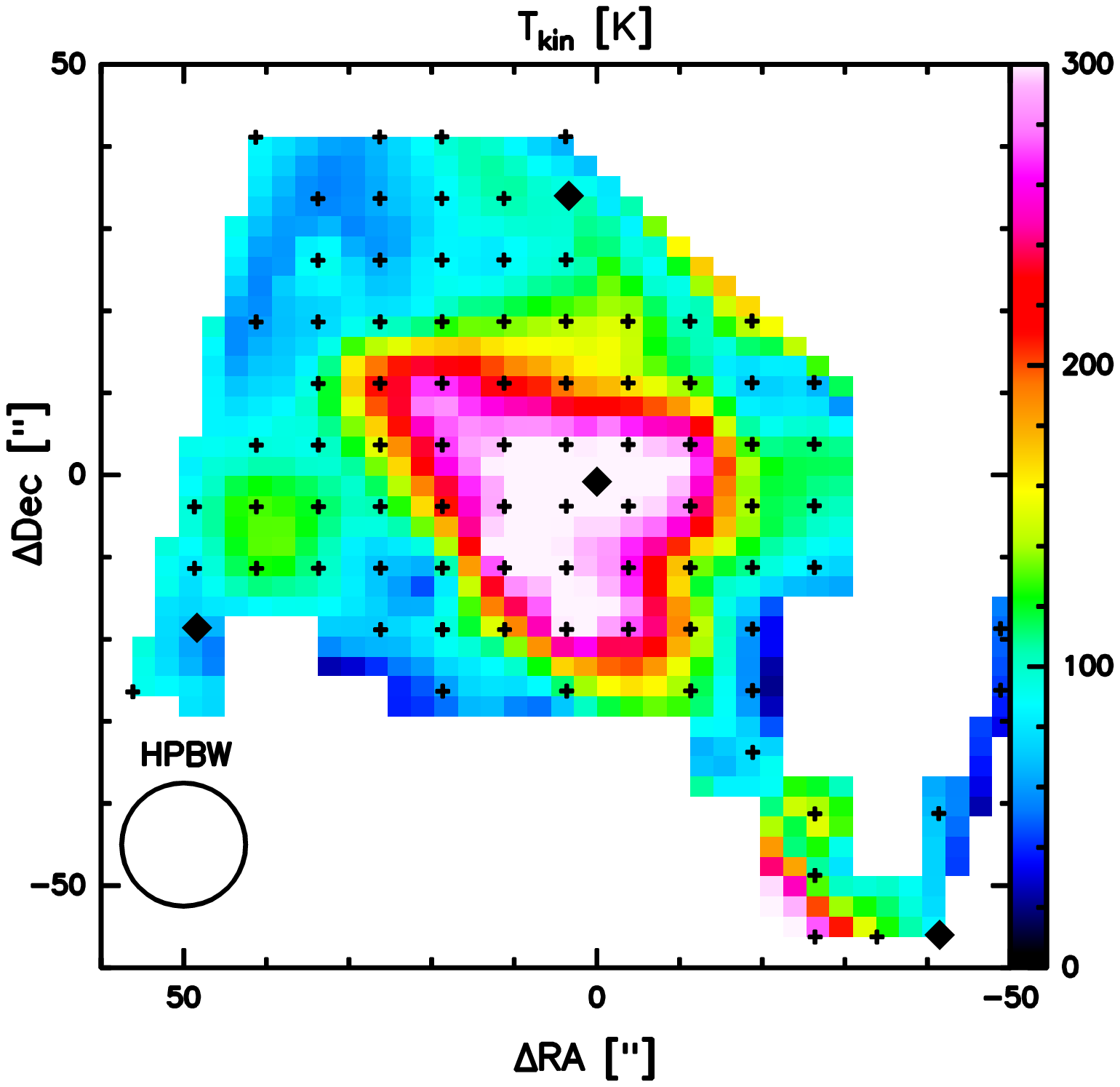}
\caption{Left panel: Ratio of $5_{33}-4_{32}$ to the $5_{15}-4_{14}$ H$_2$CO
  transition. Right panel: Kinetic temperature map of W49A determined from
  line ratio shown in the left panel.}
\label{kintemp}
\end{figure*}

The results are shown in Figure \ref{kintemp} with the main regions highlighted. We derive a kinetic temperature of 300 K toward the center, 98 K toward the Northern clump, 85 K toward the South-west clump and 68 K toward the Eastern tail, with an uncertainty of $\sim$30\%. 
In fact, there is a large, about $1'\times1'$ ($\sim3\times3$ pc) region around the center, with kinetic temperatures $\gtrsim$100 K.

\subsection{Volume- and column density estimates}
\label{sect_voldens}

\begin{figure}[!h]
\includegraphics[width=9.5 cm,trim=0.5cm 0 0 6.5cm,clip=true]{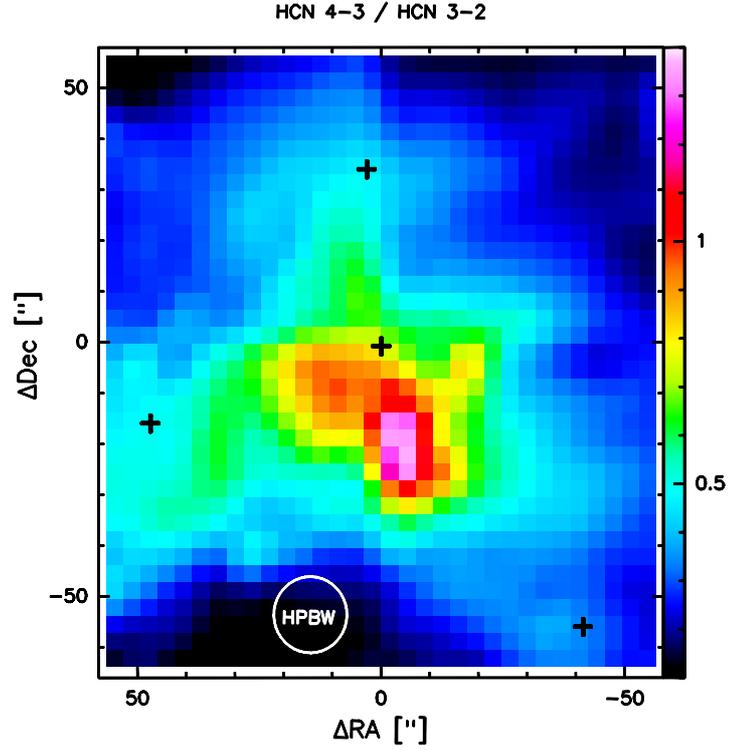}
\caption{Map of the line ratio of HCN $J$ = 4$-$3 to HCN $J$ = 3$-$2.}
\label{voldens}
\end{figure}

\begin{figure}[!h]
\includegraphics[width=9.5 cm,trim=0.5cm 0 0 8cm,clip=true]{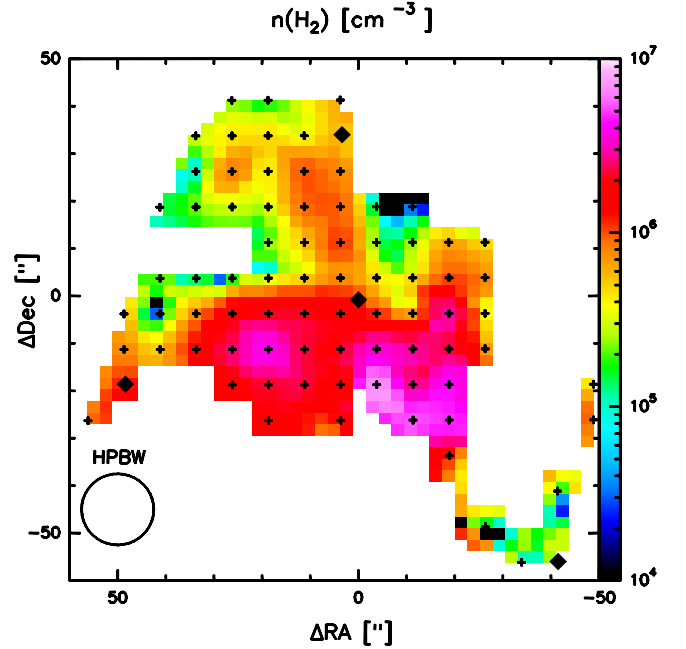}
\caption{Volume density estimates from the HCN 3$-$2 and HCN 4$-$3 lines.}
\label{dens}
\end{figure}

To estimate H$_2$ volume densities of the high temperature gas that the
H$_2$CO line ratios trace, we use line ratios of the HCN 3$-$2 (265.886 GHz)
and HCN 4$-$3 (354.505 GHz) transitions. Our calculations are based on collisional rates 
from \citet{dumouchel2010}, which have been scaled by a factor 1.37 to represent collisions with H$_2$. 
The HCN 4$-$3 map has been convolved to the resolution of the HCN 3$-$2 map (20$''$). 
Since the line ratios depend on the kinetic temperature (Figure~\ref{line_ratios_HCN}), we use the kinetic
temperatures determined from the H$_2$CO line ratios to convert the measured
HCN line ratios (Figure~\ref{voldens}) to volume densities. Therefore, even if
HCN was detected towards almost the entire SLS field, our calculations only
take into account the positions where both of the H$_2$CO transitions were
detected. The uncertainty of these density estimates is $\sim$ a factor of
two. Figure~\ref{dens} shows the estimated volume densities. Note that not all
the positions with kinetic temperature estimates are included in the density
map: this is due to the uncertainty of our method below 100 K, where the HCN
3$-$2 / 4$-$3 line ratios are rather temperature than density tracers. With a
search range in the densities between 10$^3$ and 10$^8$ cm$^{-3}$, an extended
area towards the center (approximately corresponding to the high-excitation
area revealed by the H$_2$CO line ratios) corresponds to densities as higher than 
10$^5$ cm$^{-3}$ or higher, with a maximum of $6.9\times10^6$
cm$^{-3}$. The maximum density is measured toward an offset position compared to the position we refer to as 'center'. The densities toward the central part show little less than an order of magnitude variation.

The average H$_2$ volume density corresponding to the central
$1\times1$ arcminute region is $1.6\times10^6$ cm$^{-3}$. 
The lowest densities, toward the edges of the cloud area covered by our
estimates, is a few times 10$^4$ cm$^{-3}$. However, our density
estimates are more accurate for densities between 10$^5$ and 10$^8$
cm$^{-3}$ (Figure \ref{line_ratios_HCN}). In addition to the
uncertainty of the method below $\sim$10$^5$ cm$^{-3}$, the H$_2$CO
$5_{15}-4_{14}$ and $5_{33}-4_{32}$ transitions are detected over a less
extended region, compared to the HCN 3$-$2 and 4$-$3
transitions. Estimating the physical conditions outside the area
covered by our kinetic temperature and volume density maps requires
molecular line tracers with a larger spatial extent.

Based on the kinetic temperatures and volume densities presented
above, we estimate the column densities of the two main molecular tracers used
in this paper, HCN and H$_2$CO. 
These estimates use H$_2$ volume densities of $1.8\times10^6$ cm$^{-3}$ for the center, 
$1.2\times10^6$ cm$^{-3}$ for the Eastern tail, $5.6\times10^5$ cm$^{-3}$ for the Northern 
clump and $1.3\times10^5$ cm$^{-3}$ for the South-west clump. The kinetic temperatures used for each position are listed in section \ref{sect_kintemp}.
To calculate the HCN column density toward the center, we use the H$^{13}$CN 4$-$3 (345.3 GHz) transition, 
and $^{12}$C/$^{13}$C=77 \citep{wilsonrood1994}. 
HCN peaks toward the center, with a column density of $7.3\times10^{15}~{\rm{cm}}^{-2}$.  
The HCN column densities of the Eastern tail and Northern clump regions are comparable: $N({\rm{HCN}})=7-8\times10^{13}~{\rm{cm}}^{-2}$ toward the Eastern tail and $N({\rm{HCN}})=1.5-2\times10^{14}~{\rm{cm}}^{-2}$ toward the Northern clump. N(HCN) toward the South-west clump is a factor of 4$-$5 above the column density of the Northern clump: $N({\rm{HCN}})=6.2-9.6\times10^{14}~{\rm{cm}}^{-2}$. 
$N$(H$_2$CO) is $\sim$3.2$\times10^{14}-4.5\times10^{14}~{\rm{cm}}^{-2}$ toward the center as well as the South-west clump. The Northern clump has $N$(H$_2$CO) slightly below the center and South-west clump, $7.0\times10^{13} - 1.1\times10^{14}~{\rm{cm}}^{-2}$, while toward the Eastern tail, we measure a column density of $4.9\times10^{13} - 5.0\times10^{13}~{\rm{cm}}^{-2}$. 
The two orders of magnitude difference in the H$_2$CO column density
toward the center compared to the best fit column density derived from the
rotation diagrams can be accounted for the simple assumption used in the rotation diagram method, that the level populations can be described by a single excitation temperature.

The mass of the warm and dense gas can be estimated based on the H$_2$CO column density, 
by assuming an H$_2$CO abundance. \citet{vandertak2000} derive H$_2$CO abundances in the range of $10^{-9}-10^{-10}$ for thirteen regions of massive star formation. For an 'average' column density of $10^{14}$ cm$^{-2}$ and an abundance of $10^{-9}$, the mass of the central $1'\times1'$ region with warm and dense gas is $\sim$2$\times10^4$ $M_\odot$, which becomes $\sim$2$\times10^5$ $M_\odot$ for an H$_2$CO abundance of $10^{-10}$. 
These are comparable to the mass estimates based on dust continuum measurements. 
\citet{wardthompson1990} estimate a mass of $\sim$2.4$\times10^5$ $M_\odot$ corresponding to the central $\sim$2 pc. \citet{buckleywardthompson1996} adopt a value of $\sim10^5$ $M_\odot$ as a best fit based on dust continuum as well as on CS data published by \citet{serabyn1993}.

A second estimate of the mass of dense gas can be obtained from the HCN analysis. Adopting an average number density of $1.6\times10^6$ cm$^{-3}$ over the central $1'\times1'$ region and assuming an approximately spherical region would imply a mass of $\sim$2$\times$1$0^6$ $M_\odot$. This is a factor of 10$-$100 larger than the H$_2$CO suggests indicating that the dense HCN material fills a fraction between 1\% and 10\% of the volume of this extended region.

Further evidence for a small volume filling factor is given by the best fit source sizes from the rotation diagrams toward the central position. The column densities derived from Radex are beam-averaged, while the best fit column densities from the rotation diagram correspond to clumps with sizes indicated by the best fit source size.  
Comparing the HCN column densities to the H$_2$ volume densities gives another indication of the volume filling factor. While the volume densities change between 10$^5$ and a few times 10$^6$ cm$^{-3}$ and show a rather uniform distribution, $N$(HCN) shows a 2 orders of magnitude variation (7$\times$10$^{13}-$7$\times$10$^{15}$ cm$^{-2}$) and peaks toward the center. The emission may originate in warm and dense gas in a number 
of unresolved clumps. Such a behavior was found before, by \citet{snell1984} and \citet{mundy1986} for M17, S140 and NGC 2024.

\begin{table}[!th]
\begin{minipage}[t]{\linewidth}\centering 
\caption{Results of the excitation analysis toward the center of W49A}             
\label{table:1}      
\centering                          
\begin{tabular}{p{2.4 cm} | p{1.6 cm} p{1.6 cm} p{1.6 cm}}        
  \hline\hline
&    \multicolumn{3}{c}{\textbf{I. Rotation diagrams}}\\ 
  \hline\hline                  
& \textbf{SO$_2$}& \textbf{CH$_3$OH}& \textbf{H$_2$CO}\\
\hline
\\[-1.2ex]
$T_{\mathrm{rot}}$ [K]& 156$\pm$11& 83$\pm$7& 182$\pm$84\\
$N_\mathrm{tot}$\footnote{Total beam-averaged column density calculated using the assumption that the lines are optically thin, the level populations can be characterized by a single excitation temperature and the emission fills the telescope beam.}
[10$^{15}$ cm$^{-2}$]& 
14.56$\pm$2.99& 6.58$\pm$1.38& 0.78$\pm$0.17\\
\\[-1.6ex]
Source size[$''$]& 3.1$^{+0.0}_{-0.7}$&  12.5$^{+2.5}_{-10.7}$& 2.4$^{+4.2}_{-0.3}$\\
\\[-1.6ex]
$T_{\mathrm{ex}}$ [K]& 115$^{+40}_{-15}$& $82.1^{+16.1}_{-16.1}$& 138.2$^{+111.3}_{-115.0}$\\
\\[-1.6ex]
$N_\mathrm{tot,1}$\footnote{Total source-averaged column density given by the $\chi^2$ minimization, including a correction for the optical depth and a beam dilution factor. The errors show the 1-$\sigma$ upper and lower limits of the best fit.} [$10^{15}$ cm$^{-2}$]& 
1400.0$^{+400.0}_{-430.0}$& 9.5$^{+860.5}_{-3.2}$& $43^{+1857}_{-25}$\\                                                       
\\[-1.6ex]
  \hline\hline
& \multicolumn{3}{c}{\textbf{II. Non-LTE calculation}}\\ 
  \hline\hline
  \\[-1.2ex]
  $T_\mathrm{kin}$ [K]\footnote{The uncertainty of the kinetic temperatures is $\sim$30\%.}&                  
  \multicolumn{3}{c}{300}\\
  $N$[H$_2$CO] [cm$^{-2}$]&              \multicolumn{3}{c}{3.2$-4.5\times$10$^{14}$}\\
  $N$[HCN] [cm$^{-2}$]&                  \multicolumn{3}{c}{7.3$\times$10$^{15}$}\\
  $n$ [cm$^{-3}$]\footnote{The uncertainty of the density estimates is a factor of 2.}&    
  \multicolumn{3}{c}{1.8$\times10^6$}\\
  \hline
\end{tabular}
\end{minipage}
\end{table} 

\begin{table}[!th]
\begin{minipage}[t]{\linewidth}\flushleft 
\caption{Results of the excitation analysis toward the off-center regions of W49A}             
\label{table:2}      
\centering                          
\begin{tabular}{p{2.4 cm} | p{1.6 cm} p{1.6 cm} p{1.6 cm}}        
\hline\hline
& \multicolumn{3}{c}{\textbf{I. Rotation diagrams}}\\
\hline                 
& \textbf{Northern clump}& \textbf{Eastern tail}& \textbf{South-west clump}\\
\hline
\\[-1.2ex]
$T_{\mathrm{rot}}$ [K]& 67$\pm$13& 56$\pm$3& 62$\pm$11\\
$N_\mathrm{tot}$\footnote{Total beam-averaged column density calculated using the assumption that the lines are optically thin, the level populations can be characterized by a single excitation temperature and the emission fills the telescope beam.}
[10$^{13}$ cm$^{-2}$]& 
2.26$\pm$0.48& 1.82$\pm$0.46& 1.68$\pm$0.36\\  
\\[-1.6ex]
Source size[$''$]& 14.44$^{+0.56}_{-11.2}$& 14.44$^{+0.56}_{-11.2}$& 14.44$^{+0.56}_{-10.64}$\\
\\[-1.6ex]
$T_{\mathrm{ex}}$ [K]& 67$^{+21}_{-13}$& 55$^{+14}_{-9}$& 61$^{+19}_{-10}$\\
\\[-1.6ex]
$N_\mathrm{tot,1}$\footnote{Total source-averaged column density given by the $\chi^2$ minimization, including a correction for the optical depth and a beam dilution factor. The errors show the 1-$\sigma$ upper and lower limits of the best fit.} [10$^{13}$ cm$^{-2}$]& 
2.2$^{+44.8}_{-0.2}$& 1.9$^{+39.1}_{-0.3}$& 1.7$^{+22.3}_{-0.2}$\\
\\[-1.6ex]
\hline

& \multicolumn{3}{c}{\textbf{II. Non-LTE calculation}}\\                 
\hline
& \textbf{Northern clump}& \textbf{Eastern tail}& \textbf{South-west clump}\\
\hline
\\[-1.2ex]
$T_\mathrm{kin}$ [K]\footnote{The uncertainty of the kinetic temperatures is $\sim$30\%.}& 98& 68& 85\\
$N$[H$_2$CO] [cm$^{-2}$]& 0.7$-$1.1$\times$10$^{14}$& 4.9$-$5.0$\times10^{13}$& 3.2$-$4.5$\times10^{14}$\\
$N$[HCN] [cm$^{-2}$]& 1.5$-$2$\times$10$^{14}$& 7$-$8$\times$10$^{13}$& 6.2$-$9.6$\times$10$^{14}$\\
$n$ [cm$^{-3}$]\footnote{The uncertainty of the density estimates is a factor of 2.}& 5.6$\times$10$^5$& 1.2$\times$10$^6$& 1.3$\times$10$^5$\\
\hline
\hline                                      
\end{tabular}
\end{minipage}
\end{table}

\section{Discussion}

We have detected warm ($\gtrsim$100 K) and dense
($\gtrsim$10$^5~{\rm{cm}}^{-3}$) gas extended over a $1'\times1'$
($\sim$3$\times$3 pc) field towards W49A. Based on a rotation diagram analysis
with corrections for optical depth and beam dilution, we derive excitation
temperatures of 115 K (for SO$_2$), 82 K (for CH$_3$OH), 138 K (for H$_2$CO)
towards the center.  From H$_2$CO line ratios with the same $J$ and different
$K$-states, we derive kinetic temperatures of $\sim$300 K toward the center,
68 K toward the Eastern tail, 98 K toward the Northern clump and 85 K toward
the South-West clump. Our kinetic temperature estimates are consistent with
the excitation temperatures derived from the rotation diagram method, given
that non-LTE conditions are expected, with excitation temperatures below the
kinetic temperatures. Ratios of HCN 3$-$2 / HCN 4$-$3 line intensities, using our 
kinetic temperature estimates, result in H$_2$ volume densities of $1.8\times10^6$ cm$^{-3}$ for the center, 
$1.2\times10^6$ cm$^{-3}$ for the Eastern tail, $5.6\times10^5$ cm$^{-3}$ for the Northern 
clump and $1.3\times10^5$ cm$^{-3}$ for the South-west clump.

Our kinetic temperature estimates are significantly higher than previously reported values. Based on Planck brightness temperatures derived from the CO $J$ = 7$-$6 transition, \cite{jaffe1987} estimate kinetic temperatures of $>$70 K in the centre, falling below 50 K in the outer parts of the cloud. However, these observations have a beam size of 32$''$, which suggests higher kinetic temperatures on a smaller scale, and as such, may be consistent with our results. \cite{roberts2011} use HCN/HNC line ratios compared to chemical models to estimate kinetic temperatures and derive temperatures of 100 K toward the center, 40 K toward the South-west clump, 40 K toward the Eastern tail and 75 K toward the Northern clump. One possible reason of the difference between the kinetic temperatures presented in this paper and the values estimated by \cite{roberts2011} is that \cite{roberts2011} adopted the HCN collision rates for HNC, which can result in an uncertainty, due to the fact that HCN and HNC have slightly different dipole moments (3.0 vs. 3.3 D).

Our density estimates are consistent with the estimates by \citet{welch1987}, who estimate the cloud density to be $>$3$\times10^5$ cm$^{-3}$ within the inner 1 pc and $\sim$10$^4$ cm$^{-3}$ in the outer parts of the cloud. Our estimates are also consistent with \citep{plume1997}, who find - based on CS 2$-$1, 3$-$2, 5$-$4, 14$-$13, 10$-$9 transitions - that ${n({\rm{H}}_2)}$ toward W49A varies between 5.2$\times$10$^5$ and 1.8$\times$10$^6$ cm$^{-3}$. \citet{serabyn1993} estimate densities between 2$-$6$\times$10$^6$ cm$^{-3}$ for three clumps using CS (3$-$2, 5$-$4, 7$-$6, 10$-$9) and C$^{34}$S (5$-$4, 7$-$6, 10$-$9) transitions. Our mass, volume- and column density estimates also indicate a low filling factor for the dense gas. \citep{plume1997} probe masses and densities for a number of massive star-forming regions, and find, that filling factor for the dense gas is $<$25\%.

\subsection{Possible heating sources}

The most probable mechanisms that contribute to the heating of the warm and
dense molecular gas seen toward W49A are mechanical heating (shocks produced
by stellar winds), and UV or X-ray irradiation. 
Another possible mechanism, that has been suggested for starburst galaxies, is heating by cosmic rays  \citep{bradford2003}. In the case of W49A, the only possible source of cosmic-rays above the Galactic background is the supernova remnant W49B, however, due to the high uncertainty in the distance of W49B \citep{brogantroland2001}, this remains an open question. 
Therefore, we focus on quantifying the effect of mechanical heating and irradiation by UV and X-rays for the warm and dense region revealed by our H$_2$CO and HCN data.
\paragraph{\textbf{Mechanical heating}}

The luminosity corresponding to the \emph{mechanical heating} produced by stellar winds of embedded O-type stars can be estimated from:
\begin{equation}
\label{mech}
L_{\rm{w}}=1.3\times10^{36}\left( \frac{\dot{M}}{10^{-6} M_\odot~\rm{year}^{-1}} \right) \left( \frac{v_{\rm{w}}}{2 \times 10^3~\rm{km s}^{-1}} \right)^2 \mathrm{erg~s}^{-1}
\end{equation}
where $v_{\rm{w}}=2000~\rm{km~s}^{-1}$ can be applied as the velocity of the stellar wind \citep{tielens2005}. \citet{peng2010} found evidence for two expanding shells with a size of $\sim$2.9 pc, comparable to the size of the extended high excitation region we have found. They found that a constant mass-loss rate of $\sim$1.2$\times 10^{-6}~M_\odot$ yr$^{-1}$ can sustain a wind-driven bubble with a size of $\sim$2.9 pc, which corresponds to the mass loss rate of one O-type star.
Using these values, the mechanical luminosity is $L_{\rm{w}}=1.56\times10^{36}~\mathrm{erg~s}^{-1}$ or about $405~L_\odot$. Assuming an efficiency of 1$-$10 \% \citep{loenen2008}, the mechanical heating rate is $\sim$4$-$40.5 $L_\odot$. However, this is only a lower limit of the luminosity corresponding to the mechanical heating. \cite{alveshomeier2003} estimate that $\sim$30 O-type stars belong to the central stellar cluster, which has a diameter of $\sim$6 pc. Taking this into account, the mass-loss rate, as well as the luminosity is about an order of magnitude larger than the value we derive by taking a mass-loss rate that belongs to one O-type star. This gives 40.5$-$405 $L_\odot$ as a mechanical heating luminosity.

In addition to the effect of stellar winds, protostellar outflows give a contribution to the mechanical heating. The largest outflow is connected to the most luminous water-maser emission in the Milky Way \citep{gwinn1992}. It has been characterized by \citet{scoville1986} and recently by \citet{smith2009}. 
Adopting an outflow dynamical time-scale of 10$^4$ yr, that corresponds to an average $\sim$25 km s$^{-1}$ outflow speed \citep{smith2009}, an energy of 4.4$\times$10$^{47}$ erg (\citealp{scoville1986}, scaled to a distance of 11.4 kpc), the mechanical luminosity is $L_{\rm{mech}}=1.4\times$10$^{36}$ erg s$^{-1}$ or about $\sim$400 $L_\odot$. Assuming the same efficiency as for the mechanical luminosity produced by the stellar wind, only the most powerful outflow produces a mechanical luminosity that is 10 \% of the mechanical luminosity produced by the stellar winds of an embedded cluster of $\sim$30 O-type stars. 
This estimate gives a lower limit on the mechanical luminosity produced by outflows. Additional kinetic energy injection may have been resulted by past outflows of the current O-type stars. These flows may still be dissipating energy. The existence of these flows may be probed using ALMA.

An additional source of the mechanical heating may exist in the form of shocks driven by the expansion of the (UC)H{\sc{ii}} regions into the molecular cloud. The corresponding heating rate is proportional to the kinetic energy injection rate and is given by:
\begin{equation}
\label{hiiexpansion}
G = \frac{1}{2} A \frac{\mathrm{d} \rho}{\mathrm{d} t} V^2 = \frac{1}{2} \rho A V^3
\end{equation}
where A is the surface area of the shocks, $\rho$ is the density of the medium into which the shock is moving, and $V$ is the velocity of the shock. 
The surface area of the shocks can be estimated based on \citet{depree1997}, who measure the sizes of the (UC)H{\sc{ii}} regions, based on high-resolution (0$\farcs$8) 3.6 cm data. We use an average radius of 0.05 pc to calculate the surface area of the shocks.
Based on $n(\mathrm{H}_2)=10^6$ cm$^{-3}$, a shock velocity of 10 km/s results in a heating rate of $1.6 \times 10^{37}$ for a cluster of 30 O-type stars.
For a shock velocity of 5 km/s, the mechanical heating rate of the expanding (UC)H{\sc{ii}} regions corresponding to the cluster of 30 O-type stars is $\sim2 \times 10^{36}$ erg/s - in the same order of magnitude as the mechanical heating produced by the stellar winds of the 30 O-type stars.

\paragraph{\textbf{Radiative heating by the embedded stellar cluster}}
The importance of \emph{heating by UV radiation} from young massive stars, in the form of photo-electric emission by dust grains and PAHs, can be estimated from the far-infrared luminosity. \citet{vastel2001} estimates a total FIR flux of $1.54\times10^{-6}$ erg s$^{-1}$ cm$^{-2}$, which, at the distance of W49A, is equivalent to a total IR luminosity of  $6.3\times10^6~L_\odot$. Adopting an efficiency of $10^{-3}-10^{-2}$ \citep{meijerinkspaans2005}, the heating rate by UV irradiation is on the same order of magnitude as the luminosity that can be expected from the mechanical heating.

\paragraph{\textbf{X-ray heating}}
Hard X-ray emission has been detected \citep{tsujimoto2006} toward the center of W49A, associated with two {H\sc{ii}} regions. They measure an X-ray luminosity of $3\times10^{33}$ erg s$^{-1}$ in the 3$-$8 keV band. Assuming an efficiency of 10\% \citep{meijerinkspaans2005}, the X-ray heating luminosity is on the order of 0.1 $L_\odot$, about three orders of magnitudes below the luminosity of the mechanical and UV heating.
In addition, since X-ray emission is due to point-sources, rather than an extended region, X-rays mostly act locally and seem less likely to be responsible for the extended emission region what we detected. 

\paragraph{\textbf{Gas-grain collisional heating}}

At the densities estimated using HCN transitions (10$^5$ -10$^6$ cm$^{-3}$), gas-grain collisions may give a significant contribution to the heating of the gas. 
Based on a greybody fit to the dust emission, \citet{wardthompson1990} derive a dust temperature of 50 K. They also find evidence, that a population of small, hot dust grains with $T\sim$350 K also exists, based on an excess in the near-infrared.
The heating rate of the gas per unit volume can be estimated using (Hollenbach \& McKee 1979, 1989):
\begin{eqnarray}
\label{gasgrain}
\Gamma_\mathrm{coll.}=&1.2&\times 10^{-31}n^2 \left(\frac{T_{\rm{kin}}}{1000}\right)^{1/2} \left( \frac{100~\AA}{a_{\rm{min}}} \right)^{1/2} \\
&\times & [1-0.8 \exp{(-75/T_{\rm{kin}}})] (T_{\rm{d}}-T_{\rm{kin}}) (\mathrm{erg~cm^{-3}~s^{-1}}) \nonumber
\end{eqnarray}
where the minimum grain size is set at $a_{\rm{min}}=10$ $\AA$. Using T$\sim$150 K and $n$=10$^6$ cm$^{-3}$, assuming a volume filling factor of 1-10\% for the hot gas, a gas-to-dust ratio of 100, and that the fraction of the warm ($T\sim$350 K) dust is up to 1\% of the total dust, gas-grain heating for a 3$\times$3 pc region toward the center of W49A is expected to be $\sim$10$^{34}$-10$^{35}$ erg s$^{-1}$, equivalent to $\sim$3-30 $L_\odot$.
\\

Based on these estimates, mechanical heating and UV heating seem to be the most probable heating mechanisms, while irradiation by X-rays and gas-grain collisions are less probable. The contribution of radiation to the excitation of W49A has been investigated earlier by \citet{roberts2011}, based on HCN, HNC and HCO$^+$ transitions from the SLS. By comparing the observed line intensity ratios to PDR (Photon-dominated Region) and XDR (X-ray Dominated Region) models of \citet{meijerink2007}, they find, that HCN/HCO$^+$ line ratios are consistent with PDR models, while HCN/HNC line ratios are consistent with XDR models. They interpret this result as evidence that irradiation by UV and X-ray photons plays a minor role in the excitation of W49A. However, as seen above, based on the UV heating luminosity, UV radiation cannot be ruled out to contribute significantly to the heating of W49A.

\subsection{Comparison to the cooling rate}

One of the main contributions to the total cooling in W49A is expected to occur via CO lines, therefore we derive a lower limit on the total cooling rate from the $^{12}$CO 3$-$2 (345.8 GHz) line luminosity based on data from the SLS. The CO line luminosity can be calculated from the CO 3$-$2 line intensities \citep{solomonvandenbout2005}, as
\begin{equation}
\label{l_co}
L_{\rm{CO}}~[{\rm{K~kms^{-1}~pc{^2}}}]=23.5\Omega [{\rm{arcsec}}^2] (d [{\rm{Mpc}}])^2 I_{\rm{CO}} [{\rm{K km s^{-1}}}]
\end{equation}
or alternatively, in units of $L_\odot$:
\begin{equation}
\label{l_co_lsun}
L_{\rm{CO}}~[L_\odot]=1.04\times10^{-3} S_{\rm{CO}} [{\rm{Jy}}] \Delta {\rm{V}} [{\rm{km s^{-1}}}] \nu [{\rm{GHz}}] (d [{\rm{Mpc}}])^2
\end{equation}
where $I_{\rm{CO}}=593.4$ K km/s, $T_{\rm{B}}=34.5~{\rm{K}}$, $S_{\rm{CO}}=1.2\times10^4~{\rm{Jy}}$ (using the Rayleigh-Jeans law to convert the brightness temperature to flux density) and $\Delta \rm{V}=17.2$ km/s is the average line width. These numbers correspond to a spectrum smoothed to a ${\rm{FWHM}}=1'$ beam, which covers the high-excitation region around the center.
Using these numbers, $L_{\rm{CO(3-2)}}=6.5\times10^3~{\rm{K~kms^{-1}~pc{^2}}}=9.65~L_\odot$ for the highly excited central square arcminute region. 
The fraction of the flux of the CO 3-2 transition to the total CO line flux can be estimated using Radex, adopting an H$_2$ volume density of 10$^6$ cm$^{-3}$ and a CO column density of 10$^{19}$ cm$^{-2}$. The CO column density estimate is based on the C$^{17}$O 3$-$2 transition (at $\sim$337 GHz) and a ratio of $^{16}$O/$^{17}$O$\sim$1800 \citep{wilsonrood1994}. For $T_{\rm{kin}}=100$ K, the 3$-$2 transition is expected to be 9\%, and for $T_{\rm{kin}}=300$ K, 5 \% of the total CO line flux. Using our Radex estimates of the total CO line flux:
\begin{eqnarray}
L_{\rm{CO}}&=&7.2\times10^4-1.3\times10^5~{\rm{K~kms^{-1}~pc{^2}}} \nonumber \\
           &=&107.2 - 192.9 ~L_\odot \nonumber \\
           &=&4.1 \times 10^{35} - 7.4 \times 10^{35}~{\rm{erg~s^{-1}}}, \nonumber
\end{eqnarray}
gives a lower limit on the cooling, with an accuracy of $\sim$20\%, mainly as a contribution from the typical calibration error in deriving the CO 3$-$2 line intensity. 

Based on \citet{neufeld1995}, for kinetic temperatures $>$100 K and H$_2$ volume densities $\sim$10$^6~{\rm{cm}}^{-3}$, the total cooling is up to an order of magnitude above the CO cooling, $\sim$10$^3~L_\odot$.

Our estimates of the heating rate indicate that the two most important
mechanisms to consider for W49A, are mechanical heating by the winds of
O-type stars and UV-irradiation. These two mechanisms give a heating
rate equivalent with $1000~L_\odot$, depending on the heating efficiency. 
This is of a same order of magnitude as the expected total cooling, assuming that 
the total cooling rate is a factor of 10 higher than the CO cooling rate.
This result supports that the effect of embedded O-type stars in the form of stellar winds 
and UV-irradiation can explain the high-excitation region around the center of W49A.

\subsection{Comparison to other warm and dense regions}
\paragraph{\textbf{Galactic Center}} 
In our Galaxy, a similar level of excitation to what we found toward W49A was reported for clouds near the center.
\citet{huttemeister1993} measured kinetic temperatures in 36 clouds near the Galactic center and based on NH$_3$ transitions, they have found a temperature component of $\sim$200 K, together with a cold gas and dust component. This sample, however, include no embedded stars, which could explain the highly excited gas component, and therefore, cannot be used as an analogue for W49A.
\citet{devicente1997} have detected hot molecular gas toward the Sagittarius B2 molecular cloud. The highest temperatures measured toward Sagittarius B2 are in the range between 200$-$400 K, with typical densities of $10^6-10^7$ cm$^{-3}$, similar to what we find for the center of W49A. However, these conditions correspond to hot cores with sizes in the range of 0.5$-$0.7 pc, much smaller than the region with similar physical properties in W49A. \citet{devicente1997} have also found a more extended component of warm (100$-$120 K) gas as a ring surrounding Sgr B2M and Sgr B2N with a radius of 2 pc and a thickness of 1.4 pc, with a density of about $2\times10^5~\rm{cm}^{-3}$. \citet{ferriere2007} study physical conditions in the innermost 3 kpc of the Galaxy, and find a high-temperature gas component ($T\sim150$ K), however, it corresponds to a low-density environment ($n_{\rm{H_2}}\sim10^{2.5}~\rm{cm}^{-3}$), unlike what we find for W49A.
On Galactic scales, W49A shows a uniquely extended region of high ($T\gtrsim100$ K) temperature which appears to be tracing the feedback from the young and forming stars in the region on their surrounding gas.
\paragraph{\textbf{Starbursts and active galaxies}} 

In even nearby galaxies sensitivity and angular resolution mean that the
physical conditions of the molecular gas can only be estimated on $\sim$100 pc
scales with uncertain filling factors. Nevertheless there are regions where
temperatures similar to those in W49A are measured. 
In particular, extended warm molecular gas ($>$100 K) has been
detected in a number of starburst- and Seyfert galaxies
(e.g. \cite{rigopoulou2002}).  \citet{muhle2007} have detected a warm,
$T_\mathrm{kin}\sim200$ K gas component with moderate density
($n(\mathrm{H}_2)=7\times10^3$ cm$^{-3}$) in the starburst galaxy M82,
using p-H$_2$CO line ratios, probing scales of $\sim$320
pc. \citet{ott2005} derived kinetic temperatures of 140 K and 200 K
toward two positions of the starburst galaxy NGC 253, using
NH$_3$(3,3) ATCA data on a scale of 65 pc. \citet{mauersberger2003}
have detected a warm (100$-$140 K) gas component toward the center of
three nearby starburst galaxies (NGC 253, IC 342 and Maffei 2) for
regions of 300$-$660 pc, based on NH$_3$ (1,1)-(6,6) transitions. An
even warmer component (T$>$400 K) was detected toward IC
342. \citet{papadopoulos2011}, based on a sample of 70
(Ultra-)Luminous Infrared Galaxies, most of them dominated by a warm
($T_{\rm{kin}}>100$ K) and dense ($n>10^4~{\rm{cm}}^{-3}$) gas phase
with masses of $\sim$10$^9~M_\odot$, come to the conclusion that the
main heating mechanism is rather cosmic-rays and highly supersonic
turbulence than far-UV radiation or shocks related to star-formation.
This might be one of the key differences between the extended, but
compared to scales of entire galaxies, still local phenomenon seen for
W49A, and starburst galaxies. However, in the context of the Milky
Way, W49A is a close analogue in terms of physical parameters to
regions in starburst- and active galaxies.

Another key comparison with warm ($\gtrsim$100 K) and dense ($\gtrsim$10$^5~{\mathrm{cm}}^{-3}$) regions, such as starburst galaxies, is the initial mass function (IMF). \citet{klessen2007} have found, based on hydrodynamical numerical calculations, that for regions of warm ($\sim$100 K) and dense ($\gtrsim$10$^5~{\mathrm{cm}}^{-3}$) gas, the slope of the IMF is in the range of $-$1.0 and $-$1.3, with a turn-over mass of 7 $M_\odot$. \citet{homeieralves2005} derived an IMF in W49A with a slope of $-1.70\pm0.30$ and $-1.6\pm0.3$ down to masses of $\sim$10$~M_\odot$. This result may be consistent with the calculations of \citet{klessen2007}, and if so, supports that W49A is a Galactic template for starburst galaxies. 
However, incompleteness at low luminosities and masses might explain the observed cut-off at masses of $\sim$10$~M_\odot$.

\paragraph{\textbf{Pressure as a diagnostic of starburst environments}}

Environments with increased star-formation activity, such as starburst galaxies, massive star-forming regions and clouds near the Galactic center can be characterized with their pressures, $P/k \sim nT$. Typical pressures that can be expected in the Galaxy and in the midplane of ordinary spirals are in the order of $10^4$ K cm$^{-3}$ \citep{papadopoulos2011}. 
Table \ref{table:pressure} shows the measured values of n, T and P/k for W49A, Sgr B2 and starburst galaxies. P/k values show more than an order of magnitude variation. 
Based on our derived densities and temperatures, the gas pressure in W49A is up to 5.4$\times$10$^8$ K cm$^{-3}$, similar to those measured toward SgrB2 and Arp 220. Other galactic nuclei appear to have lower pressures, which seems to be mainly the effect of a lower gas density.  
Beam dilution seems not to play a major effect for the density estimates, given the high densities measured toward Arp 220 as well as the sensitivity, which makes measurements in external galaxies biased toward the warm and dense gas. However, the estimated densities significantly depend on the tracer used, as was shown by \citet{greve2009}, who find a 10$-$100 factor difference between densities estimated from HCN compared to densities estimated from HCO$^+$. Therefore, the uncertainty in P/k is about an order of magnitude. \\ 
Pressures measured toward W49A and Sgr B2 are not entirely unique in our Galaxy. Dense gas has been detected toward a number of massive star forming regions, such as reported by \citet{plume1997}, \citet{wu2010} and \citet{mccauley2011}. However, the temperatures of those regions are only comparable to what we measure toward W49A 'locally', on sub-parsec scales, corresponding to hot cores, such as toward Orion KL \citep{mangumwootten1993}. Considering the spatial extent, pressures measured toward W49A provide a Galactic analogue of starburst galaxies.

\begin{table}[!th]
\begin{minipage}[t]{\linewidth}\centering 
\caption{Comparison of physical parameters in W49A to Galactic center clouds and starburst galaxies}             
	\label{table:pressure}      
	\centering                          
	\begin{tabular}{p{2.6 cm} | p{1.4 cm} p{1.0 cm} p{2.3 cm}}        
    \hline\hline
\\[-1.6ex]    
	\bf{Source}&            $\mathbf{n}$&              $\mathbf{T}$&              $\mathbf{P/k}$\\
	      &                 (cm$^{-3}$)&               (K)&                       (K cm$^{-3}$)\\
\\[-1.6ex]
\hline
\\[-1.6ex]                         
\bf{W49A center}&    1.8$\times$10$^6$&        300&      5.4$\times$10$^8$\\
\bf{W49A N clump}&   5.6$\times$10$^5$&        98&       5.5$\times$10$^7$\\ 
\bf{W49A E tail}&    1.2$\times$10$^6$&        68&       8.2$\times$10$^7$\\
\bf{W49A SW clump}&  1.3$\times$10$^5$&        85&       1.1$\times$10$^7$\\
\\[-1.6ex]   
\hline
\\[-1.6ex]   
\bf{Sgr B2}\footnote{\citet{devicente1997} based on CH$_3$CN J = 5$-$4, 8$-$7, 12$-$11 lines.}& 
                    2$\times$10$^5$&    100$-$120&       $2\times10^7-2.4\times10^7$\\
\bf{Henize 2-10}\footnote{\citet{bayet2004}, based on $^{12}$CO(J = 3$-$2, 4$-$3, 6$-$5, 7$-$6) and $^{13}$CO(J = 3$-$2)}&  
                    $\gtrsim$1$\times$10$^4$& 50$-$100&  $5\times10^5-10^6$\\
\bf{M82}\footnote{Median values based on three different positions \citep{naylor2010}. Obtained from multiple species radiative transfer modeling of the lines of CS, HCO$^+$, HCN, HNC, and C$^{34}$S.}&                
                    $10^{4.2}-10^{4.4}$& 130$-$160&      $10^{6.4}-10^{6.6}$\\
\bf{Arp 220}\footnote{\citet{greve2009}, based on HCN and CS lines.}&               
                    (0.3$-$1)$\times$10$^6$& 45$-$120&   $1.4\times10^7-1.2\times10^8$\\
\bf{Antennae}\footnote{\citet{zhu2003}, based on CO and  $^{13}$CO 2$-$1 and 3$-$2 lines.}&              
                    10$^5$&  90& 9$\times$10$^6$\\
\hline
\end{tabular}
\end{minipage}
\end{table} 

\section{Summary and conclusions}

We have analyzed the physical conditions towards the central $2\times2$ arcminutes region of W49A using JCMT maps with $15''$ resolution. We have reported a detection of warm ($>$100 K) and dense ($>$10$^5$ cm$^{-3}$) gas toward the center, extending up to a few parsecs. 

\begin{itemize}

\item We have characterized the excitation toward four main regions of W49A (center, Northern clump, South-West clump and Eastern tail) based on rotational diagrams of SO$_2$, CH$_3$OH and H$_2$CO, taking account the optical depth and beam dilution. High excitation and column densities, as well as significant beam dilution is seen towards the center, likely to trace embedded hot cores. The off-center positions show excitation temperatures corresponding to about half of those derived for the center, and no significant beam dilution.

\item Based on o$-$H$_2$CO line ratios, we estimate kinetic temperatures of the main regions of W49A. Toward the center, we derive a kinetic temperature of 250$-$300 K, 60$-$100 K toward the Northern clump, 80$-$150 K toward the South-west clump and 70$-$130 K toward the Eastern tail. Our estimates show an approximately $1'\times1'$ ($\sim$3.3$~\times~$3.3 pc) region with extended warm ($>$100 K) gas towards the center of W49A. 
Based on HCN 3$-$2 / HCN 4$-$3 line ratios, the high-temperature central $1'\times1'$ region has H$_2$ volume densities of $>$10$^5$ cm$^{-3}$. 
The high excitation of W49A is comparable to clouds near the center of our Galaxy and to starburst galaxies with estimates on scales of $\sim$100 pc.
 
\item The most important heating mechanisms for W49A are mechanical heating due to stellar winds of embedded, O-type stars and UV irradiation. X-ray irradiation and gas-grain collisional heating are less probable than other two mechanisms, while the effect of cosmic-rays remains an open question. However, the two main mechanisms corresponding to embedded young stellar populations provide a heating that no other mechanism is needed to contribute significantly in the heating of W49A. 
Based on earlier studies of starburst galaxies and their proposed heating mechanisms, 
W49A is a Galactic starburst analogue, in terms of excitation and physical conditions, but not in terms of the heating mechanism which yields the high excitation, due to additional heating mechanisms that have been proposed to explain the excitation if starburst- and active galaxies (such as cosmic rays or supersonic turbulence).

Our future work includes an analysis of the chemical inventory as well as kinematics of selected species, based on the entire SLS survey including data from the main survey (330$-$360 GHz, \citealp{plume2007}) and will be presented in Nagy et al. (in prep).

\end{itemize}

\begin{acknowledgements}
We thank the referees John Bally and Adam Ginsburg for the careful reading of the manuscript and for the constructive suggestions. We also thank the editor Malcolm Walmsley for additional helpful comments. The authors thank Alexandre Faure for discussion on H$_2$CO collision rates, Kuo-Song Wang for advice about population diagrams and Paul Goldsmith for discussion about the cooling rates and volume filling factors. Z. N. acknowledges Stefanie M\"{u}hle and Chris dePree for their useful comments.
\end{acknowledgements}

\bibliographystyle{aa}

\newpage

\begin{appendix}

\section{Radex line ratio plots}

\begin{figure}[!h]
\includegraphics[width=7.5 cm]{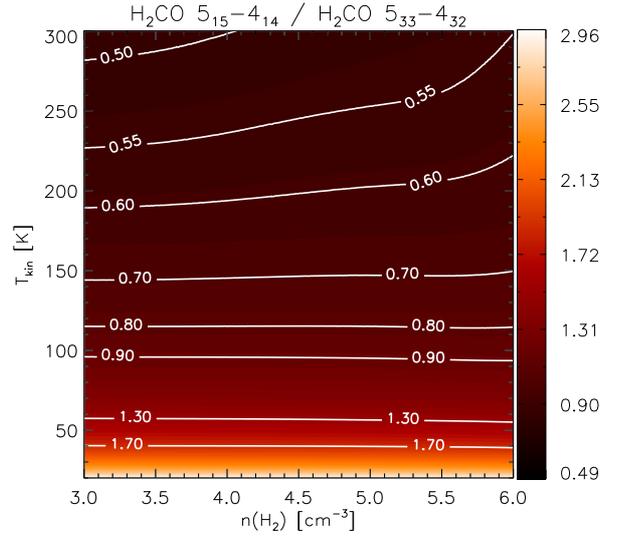}
\caption{Line ratios of o$-$H$_2$CO $5_{15}-4_{14}$ and $5_{33}-4_{32}$ as a function of kinetic temperature and H$_2$ density for FWHM=12 \kms. The intensity scale is logarithmic.}
\label{line_ratios_H2CO}
\end{figure}

\begin{figure}[!h]
\includegraphics[width=7.5 cm]{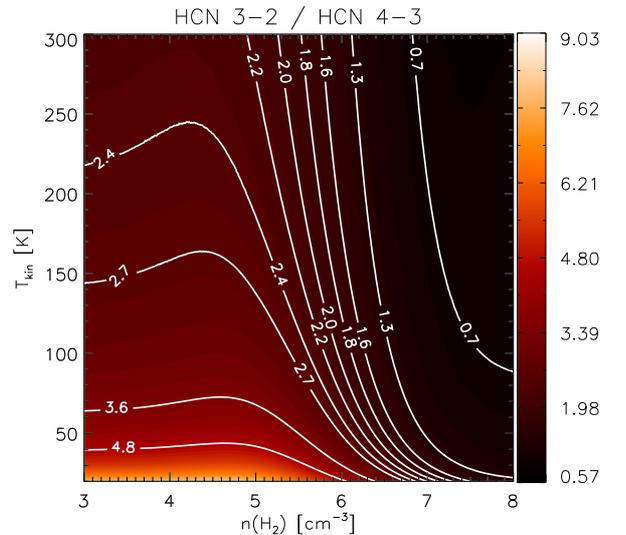}
\caption{Line ratios of HCN 3$-$2 and 4$-$3 as a function of kinetic temperature and H$_2$ density for FWHM=18 \kms.}
\label{line_ratios_HCN}
\end{figure}

In sections \ref{sect_kintemp} and \ref{sect_voldens} we compare the
observed line ratios to line ratios calculated using the non-LTE code
Radex \citep{vandertak2007}. Figure \ref{line_ratios_H2CO} shows the
the o$-$H$_2$CO $5_{15}-4_{14}$ and $5_{33}-4_{32}$ line intensity
ratios for H$_2$ volume densities between 10$^3~{\rm{cm}}^{-3}$ and
10$^6~{\rm{cm}}^{-3}$ and temperatures between 20 K and 300 K. Figure
\ref{line_ratios_HCN} shows the HCN 3$-$2 and 4$-$3 line intensity
ratios for H$_2$ column densities between 10$^3~{\rm{cm}}^{-3}$ and
10$^8~{\rm{cm}}^{-3}$ and temperatures between 20 K and 300 K. Both of
these calculations use a molecular column density of $10^{14}$
$\rm{cm}^{-2}$ and a background temperature of 2.73 K. We assume a
line width of ${\rm{FWHM}}=12$ \kms for H$_2$CO , and a line width of
18 \kms for HCN for the central $40''\times40''$, in consistence with
the observed line widths.

\end{appendix}

\end{document}